\documentclass[12pt]{iopart}

\usepackage{color}
\usepackage{graphicx}
\usepackage{cite}

\begin{document}

\title["Quest for the non-perturbative magnetic field effects in 1000-Tesla"]
{Magnetocaloric effect measurements in ultrahigh magnetic fields up to 120~T}

\author{Reon Ogawa$^{1}$, Masaki Gen$^{1*}$, Kazuyuki Matsuhira$^{2}$, and Yoshimitsu Kohama$^{1\dag}$ }

\address{$^{1}$Institute for Solid State Physics, University of Tokyo, Kashiwa, Chiba 277-8581,
Japan\\
$^{2}$Faculty of Engineering, Kyushu Institute of Technology, Kitakyushu, Fukuoka 804-8550, Japan}

\vspace{10pt}
\begin{indented}
\item[]$^*$ E-mail : gen@issp.u-tokyo.ac.jp
\item[]$^\dag$ E-mail : ykohama@issp.u-tokyo.ac.jp
\end{indented}

\begin{abstract}
We report proof-of-concept measurements of the magnetocaloric effect (MCE) in ultrahigh magnetic fields up to 120~T for the classical spin-ice compound Ho$_{2}$Ti$_{2}$O$_{7}$.
Radio-frequency resistivity measurements using an Au$_{16}$Ge$_{84}$ thin-film thermometer enable us to detect a rapid change in the sample temperature associated with a crystal-field level crossing in the high-field region in addition to a giant MCE at low fields. 
We discuss a possible delay in the temperature response and outline prospects for more precise MCE measurements in destructive pulsed fields.
\end{abstract}

\vspace{-0.1cm}
\section{Introduction}

The magnetocaloric effect (MCE) is a phenomenon in which the application of an external magnetic field to a magnetic material causes a change in its temperature.
The large MCE is observed under adiabatic conditions, where the sum of the lattice and magnetic entropies is conserved in the material.
Since the total magnetic entropy of the localized moment is given by $R \ln(2J + 1)$, where $R$ is the gas constant and $J$ is the total angular momentum, magnetic materials with large $J$, such as Gd-based compounds, can exhibit a giant MCE with temperature changes of up to several tens of kelvin \cite{1997_Pec, 2013_Kih, 2020_Miy, 2024_Tan}.
By exploiting the MCE, the concept of adiabatic demagnetization was proposed in the 1920s \cite{1926_Deb, 1927_Gia} and continues to be of great practical importance today as an environmentally friendly, gas-free magnetic refrigeration technology~\cite{2005_Bru, 2020_Kit, 2024_Zhe}.

Beyond its practical applications, the MCE serves as a powerful thermodynamic probe in condensed matter physics.
At low temperatures, the MCE tends to be pronounced due to the suppression of the lattice specific heat.
Recently, MCE measurements up to 60~T using nondestructive pulsed magnets, with pulse durations on the order of milliseconds, have been realized \cite{2013_Kih, 2020_Miy}, making it possible to map out a precise magnetic field--temperature phase diagram \cite{2014_Koh, 2017_Nom, 2019_Koh, 2019_Wei, 2022_Gen, 2023_Lee, 2024_Zha} and to elucidate quantum critical behavior \cite{2017_Bre, 2018_Wan, 2019_Gen} in low-dimensional quantum spin systems and frustrated magnets.

However, MCE measurements in ultrahigh magnetic fields exceeding 100~T have not yet been reported.
The generation of such high fields relies on destructive techniques, such as the single-turn coil (STC) method \cite{2003_Miu} and the electromagnetic flux compression method \cite{2018_Nak}, where a large discharge current gives rise to substantial electromagnetic noise and the field duration is limited to the microsecond range.
The extremely short timescale of the field duration implies that the measurements are performed under quasi-adiabatic conditions.
In this case, other heating mechanisms, such as eddy-current heating and hysteresis loss, can raise the sample temperature even more substantially than MCE.
Therefore, an accurate determination of the sample temperature under destructive pulsed fields is essential.

In this study, we performed MCE measurements in ultrahigh magnetic fields up to 120~T generated with a STC system \cite{2003_Miu}.
To detect changes in the sample temperature, we measured electrical resistivity of an Au$_{16}$Ge$_{84}$ thin-film resistive thermometer \cite{2013_Kih} using a radio-frequency (RF) impedance measurement technique \cite{2023_Shi, 2025_Chi}.
As a test sample, we selected the classical spin-ice compound Ho$_{2}$Ti$_{2}$O$_{7}$, for which a giant MCE has been reported up to 55~T using a non-destructive pulsed magnet \cite{2024_Tan}.
This compound has also been studied in magnetic fields up to 120~T by magnetization and magnetostriction measurements, revealing a crystal-field level crossing at high fields \cite{2024_Tan}.
These previous studies enable us to evaluate the validity of the present MCE results.

\section{Experimental methods}

The Ho$_{2}$Ti$_{2}$O$_{7}$ sample used here was the same single crystal as in Ref.~\cite{2024_Tan}, grown using a high-temperature halogen-lamp optical floating-zone furnace
The crystal was polished into a rectangular parallelepiped with dimensions $2\times 2\times 1.5~\mathrm{mm}^3$.
The magnetic field was applied along the [111] direction, as in Ref.~\cite{2024_Tan}.

\begin{figure}[t]
\centering
\includegraphics[width=0.95\linewidth]{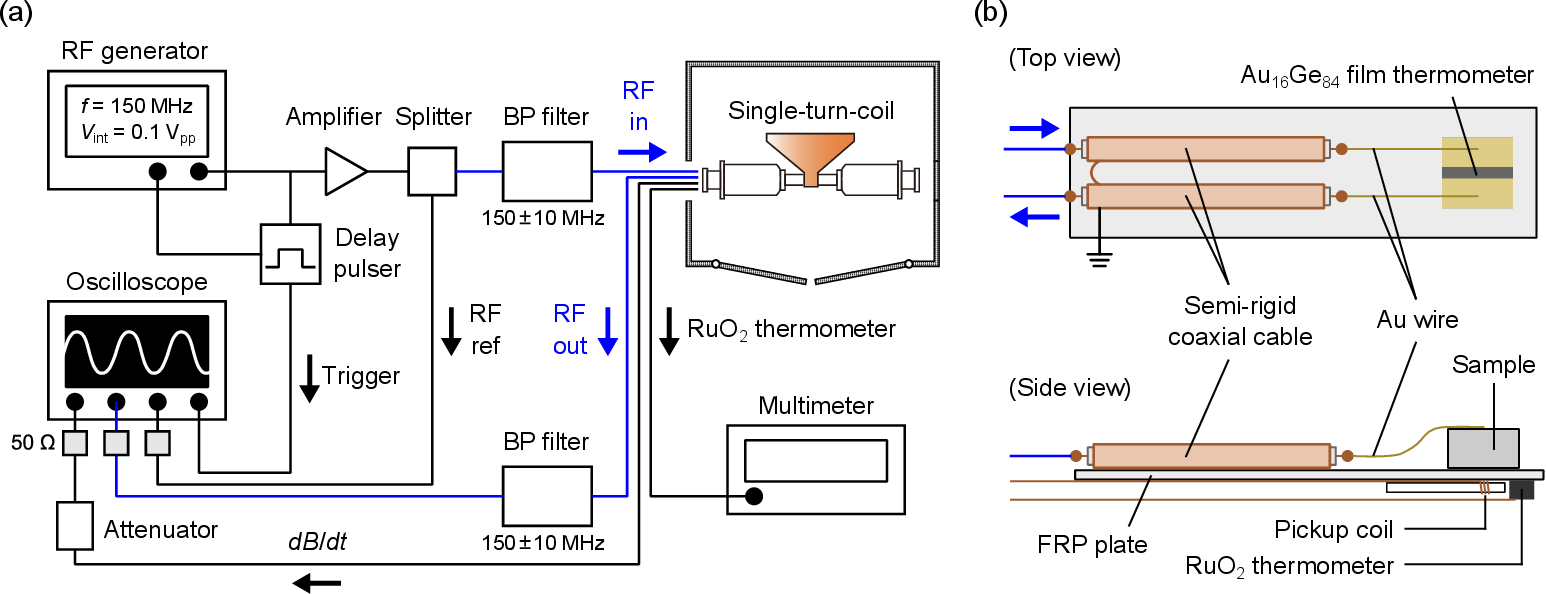}
\caption{(a) Block diagram of the MCE measurement setup in the STC system. (b) Schematic illustration of the probe and sample configuration.}
\label{Fig1}
\end{figure}

Figure~\ref{Fig1}(a) shows the block diagram of our measurement system.
For generating ultrahigh magnetic fields, we utilized a horizontal STC system at the Institute for Solid State Physics, University of Tokyo, Japan \cite{2003_Miu}.
For the RF impedance measurement, an RF sine wave at 150~MHz, generated by an RF signal generator (SRS SG382), was split into two paths using a power splitter (Mini-Circuits ZFSC-2-1-S+).
The transmission and reference signals were recorded with an oscilloscope (LeCroy 4054HD) at a sampling rate of 2.5~GS/s.
For impedance matching, 50~$\Omega$ terminators were attached to each input.
In addition, band-pass filters (R\&K, $150\pm10$~MHz) were inserted in the RF transmission circuit to suppress transient electromagnetic spikes and protect the instruments.

Figure~\ref{Fig1}(b) shows schematic illustrations of the probe and sample setting.
Two semi-rigid coaxial cables, each approximately 400~mm in length, and the sample were mounted on a fiber-reinforced plastic (FRP) plate, whereas the pickup coil for measuring the magnetic field $B$ and a RuO$_{2}$ thermometer were placed on the opposite side of the FRP plate (near the sample).
An Au$_{16}$Ge$_{84}$ thin-film thermometer was sputtered onto one face of the sample perpendicular to the (111) plane, and a pair of Au electrodes was subsequently deposited in a simple stripe geometry.
In this configuration, the self-heating of the thin-film thermometer due to eddy currents can safely be neglected.
The $R$--$T$ curve of the film thermometer shows semiconducting behavior: $R \approx 30$~$\Omega$ at 300~K, $R \approx 90$~$\Omega$ at 30~K, and $R \approx 180$~$\Omega$ at 5~K.
The core wires of the two coaxial cables were connected to each electrode by thin gold wires (30~$\mu$m diameter).
The sample was cooled down to 5~K using a liquid-$^{4}$He flow cryostat made of glass epoxy (G-10).
The resistance of the RuO$_{2}$ thermometer was monitored with a multimeter (Keithley 2000) to adjust sample temperature within the range of 5 to 30~K at zero field.

\section{Results and discussions}

\begin{figure}[t]
\centering
\includegraphics[width=0.8\linewidth]{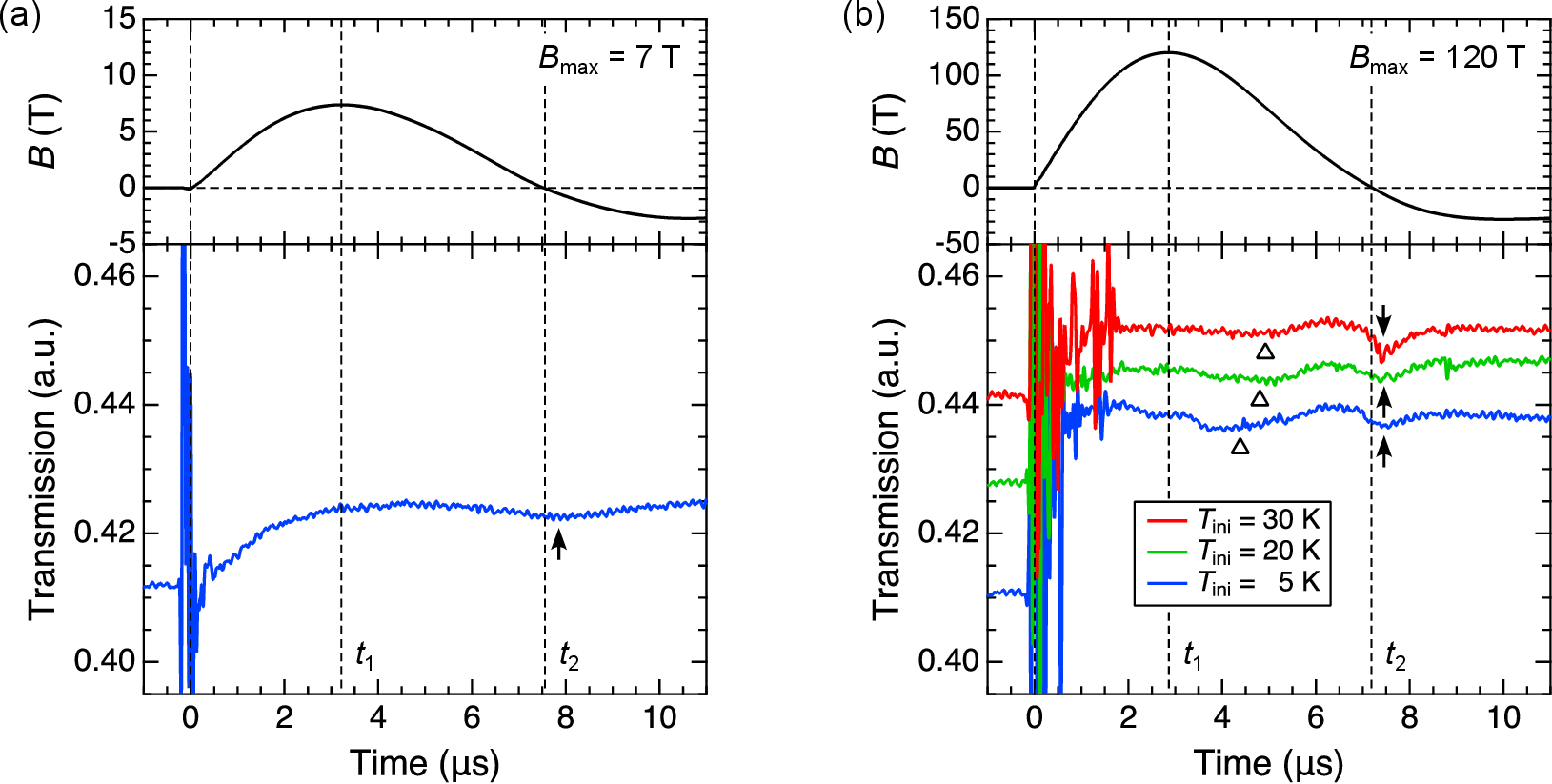}
\caption{Time profiles of the MCE measurements for (a) the $B_{\rm max} = 7$~T shot and (b) the $B_{\rm max} = 120$~T shot. Each top panel shows the magnetic-field waveform, and each bottom panel shows the detected amplitude of the RF transmission signal.}
\label{Fig2}
\end{figure}

To verify our measurement setup, we first applied a (nondestructive) pulsed magnetic field with a maximum field of $B_{\rm max} = 7$~T at an initial temperature of $T_{\rm ini} = 5$~K.
Figure~\ref{Fig2}(a) shows the time evolution of the magnetic field and the amplitude of the RF transmission signal obtained by numerical lock-in analysis, respectively.
During the field-up sweep ($0 < t < t_{1} \approx 3.2$~$\mu$s), a clear increase in the transmission amplitude is observed.
This corresponds to a decrease in the resistance and thus to an increase in the temperature of the thin-film thermometer.
Subsequently, the transmission amplitude decreases and exhibits a dip (indicated by an arrow) slightly after $t_{2} \approx 7.5$~$\mu$s, where the field reverses its sign.
The absence of a clear temperature increase during the field down-sweep suggests that the influence of eddy-current heating is not significant.
We also note that the sign of the observed resistance change is opposite to the magnetoresistance (MR) of Au$_{16}$Ge$_{84}$ \cite{2013_Kih}.
Accordingly, the present observation would reflect the intrinsic MCE in Ho$_{2}$Ti$_{2}$O$_{7}$, as reported in Ref.~\cite{2024_Tan}, where the magnetic entropy is dramatically reduced in association with the magnetic-structure change from the 2-in--2-out state to the 3-in--1-out (1-in--3-out) state.
The deviation of the timing of the dip from $t_{2}$ is $\sim$200~ns, which can be attributed to a delay of temperature response.

Figure~\ref{Fig2}(b) shows the MCE data measured in destructive pulsed fields up to 120~T at three different initial temperatures.
Although initial electrical noise results in a poor signal-to-noise ratio immediately after the discharge, the transmission signal becomes clear after $\sim$1.5~$\mu$s.
At $T_{\rm ini} = 5$~K, the temperature increase during the field-up sweep is more pronounced than that observed in the $B_{\rm max} = 7$~T shot.
For all data sets, the temperature increase tends to saturate above $\sim$100~T, consistent with the saturation of the magnetization \cite{2024_Tan}, and exhibits a dip immediately after $t_{2}$.
The reason why the dip is observed most clearly for $T_{\rm ini} = 30$~K is unclear.
In addition to the dip observed slightly after $t_{2}$, another dip appears around 50--100~T during the field-down sweep (indicated by open triangles).
This characteristic feature is likely correspond to a decrease in the sample temperature caused by an increase in magnetic entropy associated with the crystal-field level crossing expected in this field range \cite{2024_Tan}.

\begin{figure}[t]
\centering
\includegraphics[width=0.75\linewidth]{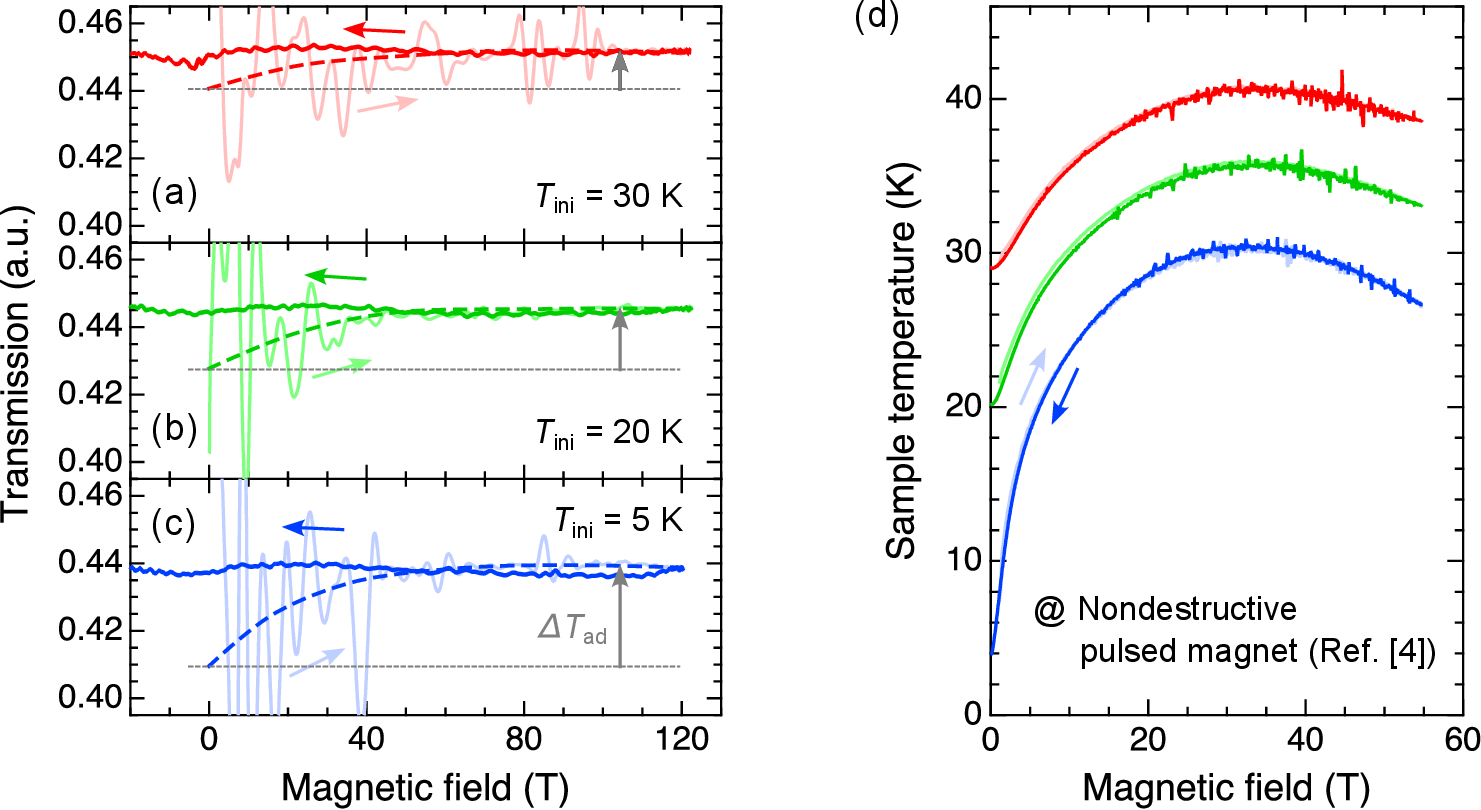}
\caption{(a)--(c) Magnetic-field dependence of the amplitude of the RF transmission signal at (a) $T_{\rm ini} = 5$~K, (b) 20~K, and (c) 30~K. The dashed lines are guides to the eye indicating the expected curves during the field-up sweep. (d) Adiabatic MCE data of Ho$_{2}$Ti$_{2}$O$_{7}$ for $B \parallel [111]$ up to 55~T obtained using a nondestructive pulsed magnet \cite{2024_Tan}.}
\label{Fig3}
\end{figure}

Figures~\ref{Fig3}(a)--\ref{Fig3}(c) show the MCE data replotted as a function of magnetic field.
Again, a larger transmission amplitude corresponds to a higher temperature.
Since the MR of the thin-film thermometer was not calibrated in the present experiments, the transmission values cannot be converted into absolute temperatures from these data alone.
We roughly estimate the adiabatic temperature increase $\Delta T_{\rm ad}$ above 100~T to be $\sim$25~K for $T_{\rm ini} = 5$~K, $\sim$15~K for $T_{\rm ini} = 20$~K, and $\sim$10~K for $T_{\rm ini} = 30$~K, given that the MR should be negligibly small above 20~K \cite{2013_Kih}.
The observed temperature increase agrees well with the MCE data up to 55~T obtained with a nondestructive pulsed magnet (36~ms duration), as shown in Fig.~\ref{Fig3}(d) \cite{2024_Tan}.
It should be noted that, in the present MCE data, the transmission amplitude does not return to its initial value when the magnetic field returns to zero, in contrast to the data measured with a nondestructive pulsed magnet.
This discrepancy can be attributed to the enhanced hysteresis loss in the microsecond pulsed-field sweep.
According to Ref.~\cite{2024_Tan}, the magnetization curve measured using a nondestructive pulsed magnet exhibit almost no hysteresis, whereas that measured using a STC system show a large hysteresis.
The present observations provide additional thermodynamic evidence for slow spin dynamics in Ho$_{2}$Ti$_{2}$O$_{7}$ arising from strong geometrical frustration.

\section{Conclusion and outlook}

In this study, we have performed MCE measurements on Ho$_{2}$Ti$_{2}$O$_{7}$ in pulsed magnetic fields up to 120~T.
In addition to the giant MCE on the order of 10~K in the low-field region, we have possibly detected a smaller temperature change of a few kelvin associated with a crystal-field level crossing in the high-field region.
These results demonstrate that the present technique, which combines a thin-film thermometer with RF resistivity measurements, is a promising approach for detecting sample-temperature changes in destructive ultrahigh magnetic fields with microsecond-duration pulses.

Finally, we point out several issues that remain to be addressed.
First, the temperature response time of the thin-film thermometer has not yet been evaluated accurately.
A previous study~\cite{2013_Koh} reported a response time of 10--100~ns, and the present MCE data suggests a similar delayed timescale of $\sim$200~ns.
Furthermore, delays arising from RF signal propagation along the cables and from phase shifts due to capacitive components in the electrical circuit are not negligible.
It is therefore necessary to establish a reliable protocol for correcting the time axis of the RF data, for example by performing MCE measurements on magnetic materials that do not exhibit hysteresis.
Another important issue is to eliminate the contribution of the MR of the thin-film thermometer.
In the nondestructive pulsed magnet, the sample temperature could be analyzed by measuring the MR of the Au$_{16}$Ge$_{84}$ film simultaneously sputtered on a nonmagnetic sample \cite{2013_Kih}.
In the STC system, however, such a calibration of the MR is not practical because of the limited sample space and the restricted number of available field shots.
It is therefore desirable to employ a thin-film thermometer made of a material that exhibits a smaller MR than that of Au$_{16}$Ge$_{84}$.

\section*{Acknowledgments}
This work was financially supported by the JSPS KAKENHI Grants-In-Aid for Scientific Research (Grants No.~24H01633 and No.~25H00600).

\end{document}